\documentclass[iop]{emulateapj}

\usepackage{graphicx}
\usepackage{epsfig}
\usepackage{epstopdf}
\usepackage[colorlinks=true,urlcolor=blue,citecolor=blue,linkcolor=blue]{hyperref}
\usepackage[T1]{fontenc}
\usepackage{amsmath}

\newcommand{\CaIIIR}{Ca~II~8542}
\newcommand{\Halpha}{H\ensuremath{\alpha}}
\newcommand{\kms}{km~s$^{-1}$}

\begin{document}

\title{Short dynamic fibrils in sunspot chromospheres}

\author{L. Rouppe van der Voort\altaffilmark{1}}
\author{J. de la Cruz Rodr\'iguez\altaffilmark{2}}

\affil{\altaffilmark{1}Institute of Theoretical Astrophysics,
  University of Oslo, %
  P.O. Box 1029 Blindern, N-0315 Oslo, Norway}
  
\affil{\altaffilmark{2}Department of Physics and Astronomy, Uppsala University, Box 516, SE-75120 Uppsala, Sweden}

\begin{abstract}
Sunspot chromospheres display vigorous oscillatory signature when
observed in chromospheric diagnostics like the strong Ca II lines and
\Halpha. 
New high-resolution sunspot observations from the Swedish 1-m Solar Telescope show 
the ubiquitous
presence of small-scale periodic 
jet-like features that move up and down. 
This phenomenon has not been described before.
Their typical width is about 0\farcs3  %
and they display clear parabolic
trajectories in space-time diagrams. The maximum extension of the top
of the jets is lowest in the umbra, a few 100 km, and
progressively longer further away from the umbra in the penumbra, with
the longest more than 1000 km. These jets resemble dynamic fibrils
found in plage regions but at smaller extensions. LTE inversion of
spectro-polarimetric \CaIIIR\ observations enabled for a comparison of
the magnetic field inclination and the properties of these short jets.
%
%
We find that the most extended of these jets also have longer periods and tend to be located in regions with more horizontal magnetic fields. 
This is a direct observational confirmation of the mechanism of
long-period waves propagating along inclined magnetic fields
into the solar chromosphere. This mechanism was identified earlier as
the driver of dynamic fibrils in plage, part of the mottles in quiet Sun, and type I spicules at the limb. 
The sunspot dynamic fibrils that we report here represent a new class of manifestation of this mechanism.
They are not the same as the transient penumbral and umbral micro-jets reported earlier.
\end{abstract}

\section{Introduction}
\label{sec:intro}

The solar chromosphere is dominated by linear structures that display vigorous dynamical evolution on a time scale from seconds to minutes. 
Off the limb, elongated features are observed as spicules with typical linear dimensions between 5--10~Mm
\citep[for a recent review see][]{2012SSRv..169..181T}.  
High-resolution {\it Hinode} observations revealed the existence of at least two types of spicules
\citep{2007PASJ...59S.655D}, 
of which the second class, or type II spicules, was found to be most abundant in the solar atmosphere.
Type I spicules are mostly found in active regions, are the minority species in quiet Sun and virtually absent in coronal holes
\citep{2012ApJ...759...18P}. 
They are characterised by clear rising and descending phases on a time scale of 140--400~s and their top describe a distinct parabolic trajectory in the space-time domain. 
Their properties agree well with what was measured on the solar disk for dynamic fibrils in active region plage by 
\citet{2007ApJ...655..624D} 
and for some mottles in quiet Sun by
\citet{2007ApJ...660L.169R}. 

In strong plage regions, dynamic fibrils display a striking quasi-periodic 
up- and down-motion with dynamic fibrils re-occurring at roughly the same location over practically the full duration of observational datasets. 
Often, dynamic fibrils show group behaviour with neighbouring structures moving in harmony, or with a small phase delay.
Numerical simulations demonstrated that this wave-like behaviour is driven by slow-mode magneto-acoustic shocks that result from $p$-mode waves leaking from the photosphere into higher atmospheric layers
\citep{
1982SoPh...75...99S, 
1982SoPh...78..333S, 
2006ApJ...647L..73H, 
2007ApJ...655..624D, 
2007ApJ...666.1277H, 
2011ApJ...743..142H}. 
\citet{2004Natur.430..536D} showed that the 5-minute $p$-mode oscillations, that earlier were considered to be evanescent above the photosphere, can propagate progressively more efficient along increasingly more horizontal magnetic fields.
When these long-period waves propagate to higher atmospheric regions with lower density, they turn into shocks that drive the observed dynamic fibrils. 
Such change in cutoff frequency for wave propagation along slanted fields had earlier been described by
\citet{1973SoPh...30...47M} 
and
\citet{1977A&A....55..239B} 
but was not appreciated in the subsequent literature except for
\citet{1990LNP...367..211S}. 

In this paper we study comparable phenomena in sunspots.
Sunspot atmospheres are also clearly dominated by waves and shocks: chromospheric diagnostics like \Halpha\ and the strong Ca~II lines show sunspots and pores oscillating in intensity and Doppler signal coherently over large fractions of their surface areas. 
This captivating phenomenon is particularly striking in the Ca~II~H and K lines where the oscillations are manifested as strong intensity variations or umbral flashes and running penumbral waves 
\citep[see e.g., the movies accompanying][]{2003A&A...403..277R, 
2007PASJ...59S.631N}. 
There has been observational evidence that waves in the chromosphere of sunspots are driven from below
\citep[see e.g., the review by][]{1992sto..work..261L}. 
\citet{2010ApJ...722..888B} 
perform non-LTE radiation hydrodynamic simulations of the propagation of acoustic waves in sunspot umbrae. 
They conclude that Ca~II~H \& K umbral flashes result from increased emission of the local solar material during the passage of waves originating in the photosphere and steepening to shocks in the chromosphere. 
While umbral flashes are a relatively large-scale phenomenon with emission patches of several arc-seconds across, there have been reports that they harbour fine-structure. 
This was concluded from polarimetric inversions 
\citep{2000ApJ...544.1141S, 
2000Sci...288.1396S} 
and direct imaging 
\citep{2009ApJ...696.1683S, 
2013A&A...557A...5H}. 
In this paper we present high-resolution \Halpha\ and \CaIIIR\ observations that reveal 
for the first time
that sunspot chromospheres are ubiquitously filled with periodic jets that resemble dynamic fibrils 
in their physics but at significantly smaller extent.

\section{Observations and Data Processing}
\label{sec:obs}

Two observational datasets obtained with the Swedish 1-m Solar Telescope
\citep[SST,][]{2003SPIE.4853..341S} 
on La Palma are analysed: one from the 
Solar Optical Universal Polarimeter \citep[SOUP,][]{title81SOUP} 
and one from the CRisp Imaging SpectroPolarimeter 
\citep[CRISP,][]{2008ApJ...689L..69S}. 
The CRISP dataset is the main dataset and will be discussed first.

CRISP is a Fabry-P{\'e}rot tunable filter instrument that is capable of fast wavelength switching ($<50$~ms) so that densely sampled spectral profiles can be obtained in relatively short time. 
The time series analyzed here was obtained on 2011 May 4 between 08:30 and 09:22~UT. 
The target sunspot (AR11204) was located at heliocentric coordinates $(x,y)=(-336\arcsec,332\arcsec)$ and observing angle $\theta=30\degr$ ($\mu=0.87$).

The CRISP observing program covered both the \Halpha\ and \CaIIIR\ spectral lines. 
\Halpha\ was sampled at two line positions, at line center and in the red wing at $+1300$~m\AA. 
\CaIIIR\ was sampled at 14 line positions in spectro-polarimetric mode so that profiles were obtained for the four Stokes parameters $I, Q, U,$ and $V$. The sampling was 9 positions between $\Delta\lambda=\pm300$~m\AA\ at 75~m\AA\ steps, plus $\Delta\lambda=\pm400, \pm860, $ and $+3100$~m\AA\ from line center.
The time to complete the acquisition of the two lines was less than 16~s, the 52~min time series comprises 193 time steps. 

High spatial resolution close to the diffraction limit of the telescope was achieved with the aid of the adaptive optics system 
\citep{2003SPIE.4853..370S},  
and image restoration using Multi-Object Multi-Frame Blind Deconvolution 
\citep[MOMFBD,][]{2005SoPh..228..191V}. 
The processing of the CRISP data was based on the various procedures developed by
\citet{2011A&A...534A..45S}, 
\citet{2012A&A...548A.114H}, 
\citet{2008A&A...489..429V}, 
and \citet{2010selbing}.
For a detailed description of the treatment of this dataset, in particular the polarimetric \CaIIIR\ data, we refer to 
\citet{2013arXiv1304.0752D}, 
who analyzed the same data in the context of umbral flashes.

After image restoration of the individual line scans and time steps, the time series were corrected for field rotation, and residual offsets and rubber-sheet deformations following 
\citet{1994shine} 
using the wide-band images as reference. 
The \Halpha\ time series were then aligned to the \CaIIIR\ time series to account for small differences in the optical path for the two lines. 
The resulting effective field-of-view of the time series was 53\arcsec$\times$52\arcsec\ with an image scale of 0\farcs059 pixel$^{-1}$.

The SOUP dataset was obtained on 2005 October 4 and was analysed by 
\citet{2007ApJ...655..624D}. 
The time series is of 78 minute duration, has 1~s cadence and centred on the \Halpha\ line core only. The field-of-view was centred on one of the larger spots (of about 20\arcsec\ diameter) in AR 10813 at $\mu=0.78$. 
Full characterisation of the instrumentation and data is presented in 
\citet{2007ApJ...655..624D}. 
For the remainder of the paper we focus mainly on the CRISP dataset, when we analyse the SOUP data this is explicitly mentioned.

\begin{figure*}[!t]
\begin{center}
\includegraphics[width=\textwidth]{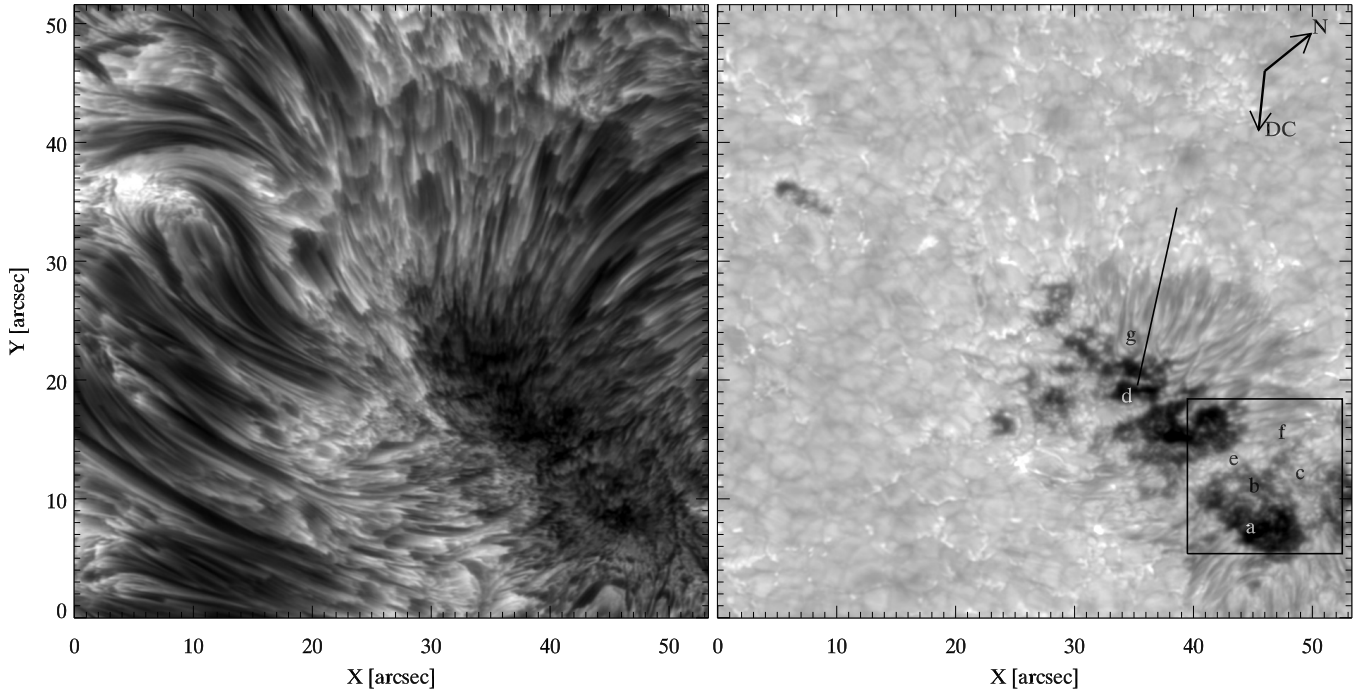}
\caption{Overview of the observed area. The left panel shows \Halpha\ line core, the right panel the co-temporal wideband image of the FWHM 4.9 CRISP prefilter centered on \Halpha. The square in the lower right outlines the region shown in more detail in Figure~\ref{fig:zoomin}. The solid line marks the position of the $xt$-slice shown in Figure~\ref{fig:xt}. Letters a--g mark locations for the $\lambda t$-diagrams in Figure~\ref{fig:lamt}. The arrows point in the direction of solar North and disk center (DC). The temporal evolution of the time series can be viewed as an animation of this figure in the on-line material.}
\label{fig:fov}
\end{center}
\end{figure*}

\subsection{Description of Field of View}
\label{sec:fov}

One corner of the field of view is dominated by the large leading sunspot of AR11204, see Figure~\ref{fig:fov}. The wideband image shows a sunspot with a double umbra that is separated by a strong light bridge. 
Some regions of the umbra have dense concentrations of umbral dots or light bridge-like structures. 
The umbra is partly surrounded by penumbra, a large section that is facing the trailing, opposite polarity of the active region does not have penumbra.
The small pore at $(X,Y)=(7\arcsec,35\arcsec)$ 
has opposite magnetic polarity and is at the front edge of the trailing plage region that is mostly outside the field of view. 

The \Halpha\ image in Figure~\ref{fig:fov} shows an active region chromosphere where the left half of the image is dominated by long ($>20\arcsec$), apparently almost horizontal fibrils that connect the two polarities in the active region. 
Similar long fibrils can be found in the upper right part of the sunspot and seem to be rooted in the outer part of the penumbra. 
Shorter fibrils are found outside the sunspot in the region at the lower left of the sunspot 
(outlined by $X=20\arcsec$--$35\arcsec$ and $Y=0\arcsec$--$20\arcsec$), 
and above the sunspot right from the middle of the image (outlined by $X=25\arcsec$--$40\arcsec$ and $Y=25\arcsec$--$52\arcsec$).
These shorter fibrils are more dynamic then the longer fibrils and periodically rise and fall in a matter of minutes. 
This is can be clearly seen in the \Halpha\ movie that is part of the on-line animation of Figure~\ref{fig:fov}. 
These are dynamic fibrils, they were described in detail by 
\citet{2006ApJ...647L..73H} 
and \citet{2007ApJ...655..624D}. 
For the remainder of this paper, we refer to these two papers as VH06 and BDP07.

In the umbra, the \Halpha\ image shows a lot of fine detail in the form of dark specks against a brighter background. In the inner-penumbra, the darker specks are more elongated. 
These darks specks are periodically varying in extension and linear position in the \Halpha\ movie and appear to be superposed on the large scale pattern of intensity oscillations that is associated with the well-known chromospheric sunspot oscillations and running penumbral waves. 
Zooming-in on the umbra and inner-penumbra (best in the on-line animation associated with Figure~\ref{fig:zoomin}), it is clear that the dark specks periodically move up and down, on a smaller scale but in a similar way as the dynamic fibrils in the plage regions outside the sunspot. 
These small-scale sunspot dynamic fibrils are the main subject of analysis in this paper.

\subsection{Measurement Methods}
\label{sec:methods}

The multi-dimensional CRISP dataset (with multiple diagnostics \Halpha\ and \CaIIIR\ and associated wideband channels, spatial dimensions $X$ and $Y$, temporal dimension $t$ and spectral dimension $\lambda$) can be effectively explored with the Crisp Spectral Explorer  
\citep[CRISPEX,][]{2012ApJ...750...22V}. 
This is a widget-based IDL tool that allows quick access to practically any cut through the dataset.
CRISPEX offers video-player functionality for visualization of the time series of the two spectral lines side by side and functionality for zooming into regions of interest and quickly switching between spectral line positions. 
Quick inspection of the spectral evolution at different locations is possible by real-time visualization of the $\lambda t$-diagram of the pixel under the mouse cursor.
Space-time diagrams or $xt$-diagrams can be quickly extracted from the data along any spatial path (linear or curved) and the associated data can be stored separately for more detailed analysis. 
The versatile functionality of CRISPEX was an effective way to confirm the nature of the plage fibrils as dynamic fibrils in the context of the work by 
VH06 
and BDP07, 
and to establish the similarity of the small-scale sunspot fibrils to the dynamic fibrils. 

For measurement of physical properties of the sunspot fibrils, we follow the same methods as 
VH06, BDP07, 
\citet{2007ApJ...660L.169R} 
and \citet{2008ApJ...673.1201L}. 
Space-time diagrams were extracted along linear paths aligned with the main axis of individual fibrils. 
Just as for dynamic fibrils in plage, the tops of the fibrils follow a path that is well approximated by a parabola.
The parabolic fit to the top of fibril in the $xt$-diagram is then used to determine the velocity, deceleration, duration and maximum height of the fibril. 

To investigate the relation of the magnetic field vector with the fibril properties, we inverted the Stokes \CaIIIR\ data of a 34\arcsec$\times$27\arcsec\ region to determine 
magnetic field vectors. 
We used the spectropolarimetric inversion code Nicole 
\citep{2000socas-navarro} 
for LTE inversions of 286$\times$225 pixels 
(every second pixel along the $x$ and $y$ dimension). 
\citet{2012delacruz} suggested 
the reliability of LTE inversions for chromospheric magnetic field measurements, which we conveniently use in this work to retrieve the magnetic field vector, although Nicole is primarily a non-LTE inversion code.

More details on the inversions of the \CaIIIR\ data can be found in 
\citet{2013arXiv1304.0752D}. 
In that study, LTE inversions were performed for the full time series of a small patch that covered the umbra and inner penumbra in a region with strong Stokes $Q$ and $U$ signal. 
It was shown that the magnetic field inclination for this patch was very stable throughout the time series which justifies our generalization of using one time step as a measure of the inclination for the whole time series.
The inversions provide information on the magnetic field vector at a height that is well above the level that is probed by standard photospheric polarimetric diagnostics. 
The peak in polarimetric sensitivity for the \CaIIIR\ line is around $\pm200$~m\AA, which probes the low chromosphere or an approximate height of 800~km in quiet Sun 
\citep[see][]{2012delacruz}. 

\section{Dynamic Fibrils in the sunspot}
\label{sec:results}

\begin{figure*}[!t]
\begin{center}
\includegraphics[width=\textwidth]{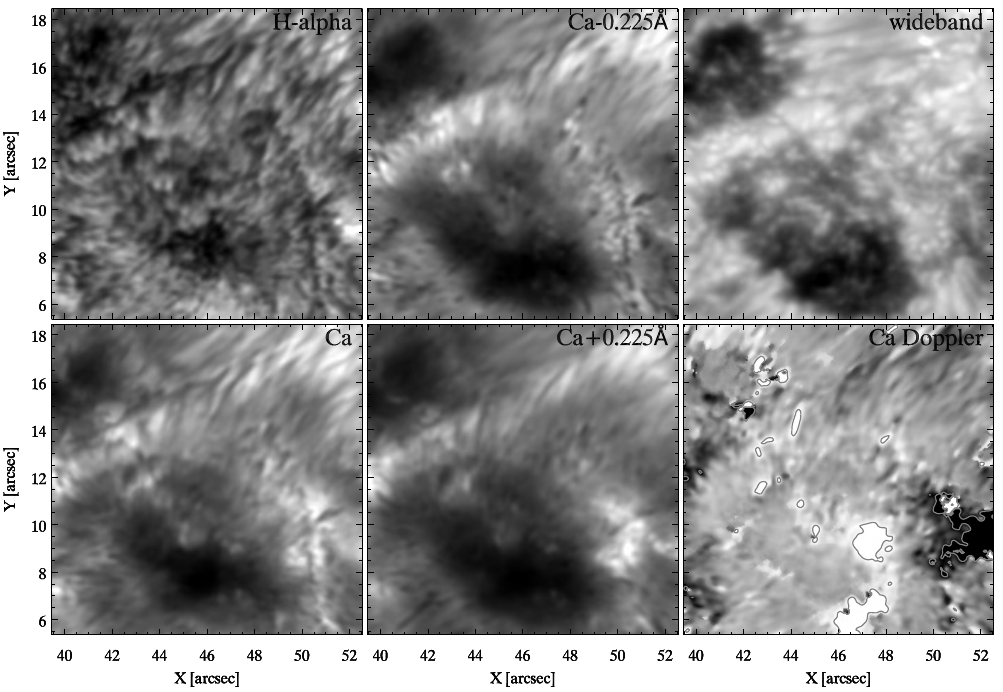}
\caption{Zoom-in on part of the sunspot (region marked with a square in Figure~\ref{fig:fov}). Upper left: \Halpha\ line center. Lower left: map of \CaIIIR\ line minimum intensity. Upper middle: \CaIIIR\ blue wing at $-0.225$~\AA. Lower middle: \CaIIIR\ red wing at $+0.225$~\AA. Upper right: photosphere in \Halpha\ wideband. Lower right: map of \CaIIIR\ Dopplershift, scaled between $\pm5$~\kms, grey contours mark regions where the parabolic fit did not yield a reasonable Doppler measurement. An animation of this figure is available as on-line material.}
\label{fig:zoomin}
\end{center}
\end{figure*}

Figure~\ref{fig:zoomin} shows a zoom-in on part of the sunspot for different diagnostics. 
The \Halpha\ image is an image with CRISP tuned to the nominal line center. 
The Ca image below is a map of the minimum intensity in the line per pixel (so it is not the \CaIIIR\  image of the CRISP nominal line center tuning).  
The reason for showing the line minimum intensity accounting for wavelength shift is that the \CaIIIR\ line is much more sensitive to Dopplershifts that \Halpha\ and the amplitude of the Dopplershifts associated with the dynamic fibril motion is large enough that one cannot cover the full evolution of the fibril by just tuning to one fixed wavelength. An image of the \CaIIIR\ line minimum intensity shows the dynamic fibrils much clearer. 
The middle panels show the blue and red wing of \CaIIIR\ at a corresponding Doppler offset of 8~\kms, 
the Ca Doppler image is a map of the shift of the \CaIIIR\ line measured from a parabolic fit to the central 5 line positions. 
This simple method is only relevant for simple absorption profiles, not for complicated line profiles with emission reversals as for example umbral flashes \citep[see ][]{2013arXiv1304.0752D}  
or flat bottom (``raised core'') profiles
\citep{2013ApJ...764L..11D}. 
The wideband image serves as photospheric reference.

While the small-scale fibrils are visible in the \Halpha\ and \CaIIIR\ panels of Figure~\ref{fig:zoomin}, we strongly encourage the reader to view the accompanying animation that shows the full time series for all panels. 
In the \Halpha\ and Ca movies, the periodic linear up and down motion of the small fibrils is very clear. 
Blinking of the \Halpha\ and \CaIIIR\ line minimum intensity images shows a close correspondence of the the small fibrils in both diagnostics with a small offset of \Halpha\ being slightly taller, or higher, than Ca, suggesting a slightly higher formation height for \Halpha\ than \CaIIIR.
A striking difference between the two diagnostics is that in \Halpha\ the pattern of small dark specks on a brighter background is visible everywhere in the sunspot. 
In \CaIIIR\ on the other hand there are patches dominated by strong intensity fluctuations caused by umbral flashes where the small fibrils are hardly or not at all discernible. 
The still Ca image in Figure~\ref{fig:zoomin} shows such a region as a dark patch between $X=44\arcsec$--$47\arcsec$ and $Y=6\arcsec$--$9\arcsec$. At this instant, the region is in a quiescent state between flashes. 

The small fibrils are clearly visible in the blue and red wings in the middle panels of Figure~\ref{fig:zoomin} and the movie vividly displays the effect of Dopplershifts in this diagnostic: in the blue wing, the fibrils are exclusively visible as growing and up-moving (and therefore blue-shifted), while in the red wing the dark fibrils are exclusively shrinking and down-moving. 
The Ca Doppler image shows both dark (blue-shifted) and white (red-shifted) patches and in the movie one can see both up-moving blue-shifted (black) fibrils and down-moving red-shifted (white) fibrils.
The up-moving blue-shifted fibrils appear to dominate the Doppler movie, this might be a visual impression (possibly the eye is more susceptible for moving dark features) combined with the possibility that the simple method of measuring the Dopplershift from a parabolic fit to the line is less effective for the down-moving phase.

The typical width of the fibrils is found to be around 0\farcs3 (200~km). 
We sometimes see faint examples with widths down to the diffraction limit of the telescope ($\lambda/D=0\farcs14$ at $\lambda=656.3$~nm).
Wider examples of order 0\farcs5 sometimes give the impression of harbouring sub-structure as if they are pairs of closely aligned fibrils.

\begin{figure}[!t]
\begin{center}
\includegraphics[width=\columnwidth]{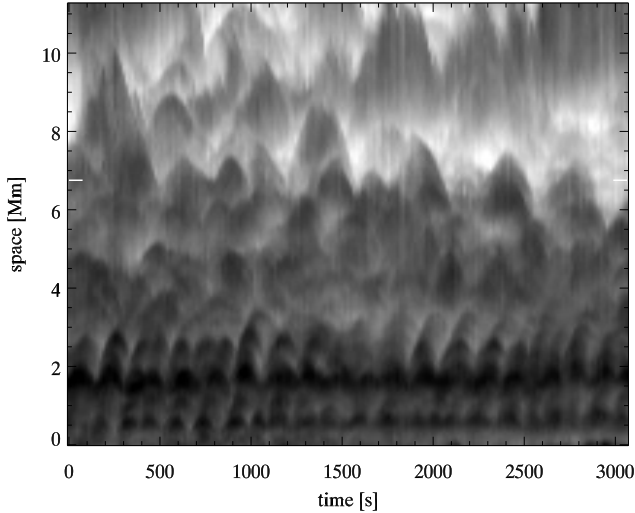}
\caption{\Halpha\ $xt$-diagram along a line that starts in the umbra (at 0) and crosses through the penumbra parallel to the penumbral filaments (see line marked in Figure~\ref{fig:fov}). The (photospheric) outer penumbral boundary is at 6.8~Mm and is marked with a thin white line.}
\label{fig:xt}
\end{center}
\end{figure}
%

\begin{figure}[!t]
\begin{center}
\includegraphics[width=\columnwidth]{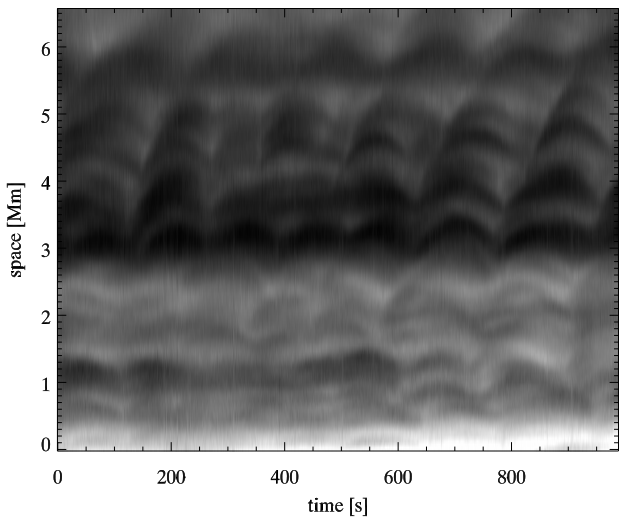}
\caption{\Halpha\ $xt$-diagram from the 2005 October 4 SOUP data with 1~s cadence. The slit covered almost completely the umbra (in the bottom of the diagram), the full (but narrow) penumbra, and a small piece of the surroundings (from about 5.5~Mm). }
\label{fig:xt04Oct2005}
\end{center}
\end{figure}

The parabolic nature of the trajectory of the top of the dynamic fibrils in the sunspot is clearly visualized in Figure~\ref{fig:xt}. 
An $xt$-diagram is shown for a linear path that crosses from the umbra through the penumbra and into surrounding plage (see location of the path marked in Figure~\ref{fig:fov}). 
%
The different extensions of the parabolas depending on the position along the path or location in the sunspot is striking: from very small in the umbra in the bottom of the diagram, to progressively larger parabolas in the outer penumbra and beyond the sunspot. 
The umbral parabolas also have clearly shorter duration than the parabolas in the outer penumbra and surrounding plage. 
Another notable regional difference in the $xt$-diagram is the regularity of the appearance of parabolas in the umbra and inner penumbra, and more irregular appearance of parabolas further out. 
In the region $x\sim2$--3~Mm, different parabolas appear progressively later or with a phase delay. This is particularly clear for $t=1900$--2500~s. This part of the diagram is crossing the inner penumbra where running penumbral waves are observed. 

The sunspot dynamic fibrils were highly 
resolved in the temporal domain in the 1~s cadence SOUP dataset. 
Various of the above-mentioned aspects of the $xt$-diagram are also prominent in the SOUP $xt$-diagram in Figure~\ref{fig:xt04Oct2005}: the parabolic trajectories of the dynamic fibril tops, varying extensions (low in the umbra, progressively larger in the outer umbra and penumbra), and regularity in recurring dynamic fibrils. 
In the region $x\sim3$--5~Mm, we observe a similar phase delay in the occurrence of dynamic fibrils as in Figure~\ref{fig:xt}.

\begin{figure*}[!t]
\begin{center}
\includegraphics[bb = 0 23 480 123, clip, width=\textwidth]{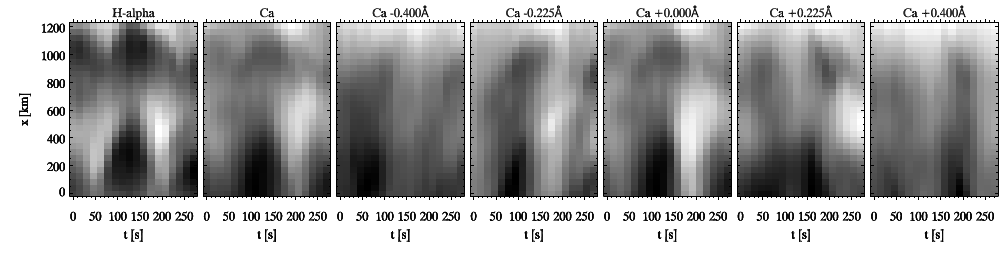}
\includegraphics[bb = 0 23 480 113, clip, width=\textwidth]{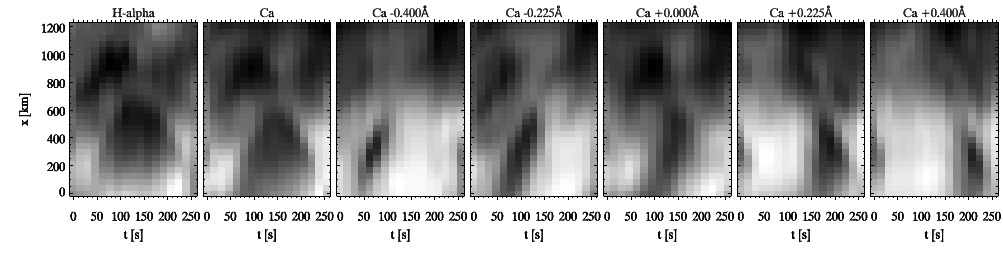}
\includegraphics[bb = 0 0 480 113, clip, width=\textwidth]{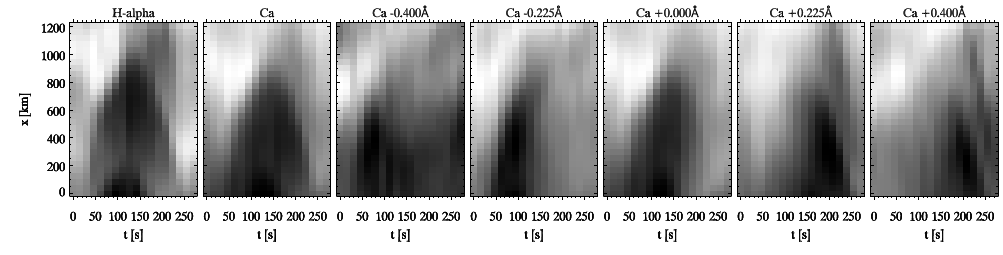}
\caption{$xt$-diagrams for three different DFs in various diagnostics marked at the top of the columns. Top row: umbral DF, middle row: DF in light bridge, bottom row: penumbral DF. The column marked ``Ca'' is \CaIIIR\ line minimum intensity, ``Ca $+0.000$~\AA" is at the nominal line center at 8542~\AA. }
\label{fig:dfdetails}
\end{center}
\end{figure*}

Detailed $xt$-diagrams for three different dynamic fibrils are shown for various diagnostics in Figure~\ref{fig:dfdetails}: \Halpha\ line core, \CaIIIR\ minimum intensity, and 5 different line positions in \CaIIIR. 
The shortest (both in length and duration) dynamic fibril is in the umbra (top row), the dynamic fibril in the light bridge is slightly taller and of longer duration (middle row), and the penumbral dynamic fibril is tallest (bottom row). 
From comparison of the left two columns (and the fifth column ``Ca +0.000\AA'' at nominal line center) we see that the dynamic fibril in \Halpha\ is slightly above \CaIIIR. 
It is interesting to note how the different phases in the life of a dynamic fibril can be traced through the wings of \CaIIIR: the dynamic fibril first appears in the blue wing at $-0.400$~\AA\ and $-0.225$~\AA, and has disappeared again in the blue wing when it reaches maximum height at line center.
After maximum height, the down moving phase can be followed in the red wing positions. 
%

\begin{figure*}[!t]
\begin{center}
\includegraphics[width=\textwidth]{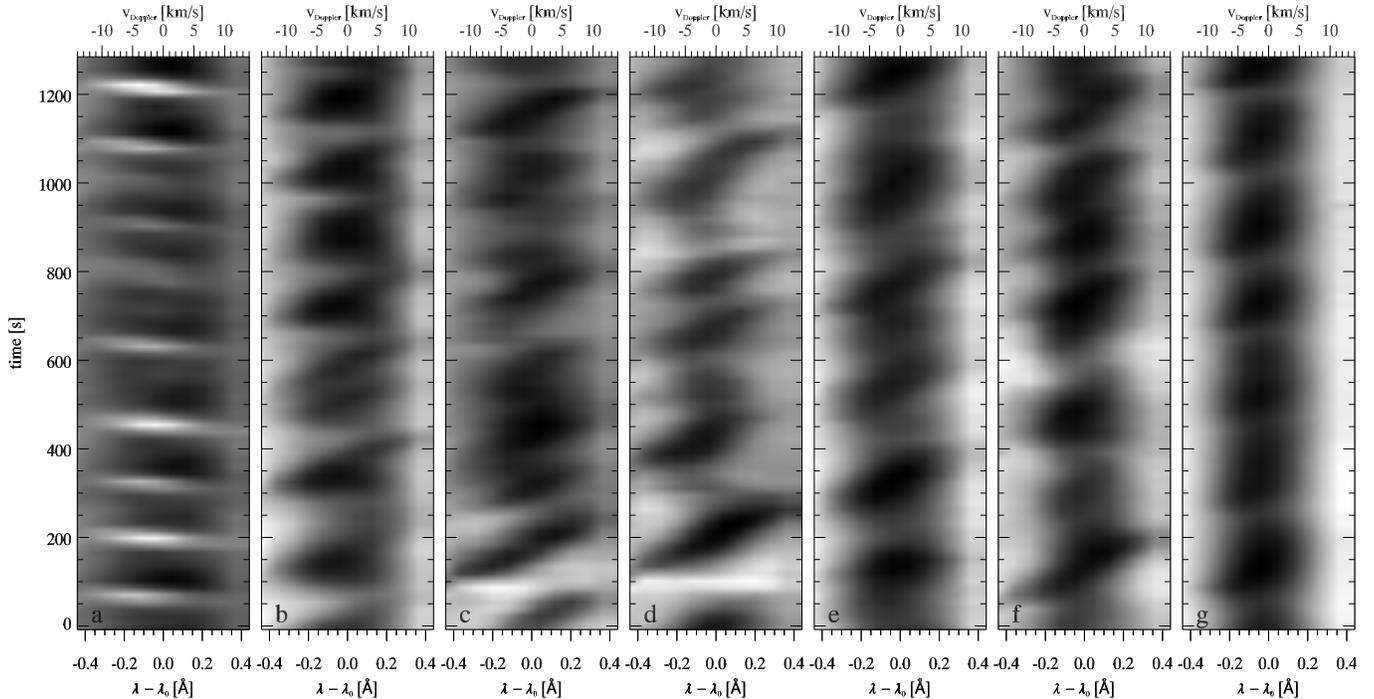}
\caption{\CaIIIR\ $\lambda t$-diagrams for various locations in the sunspot. a: in dark umbral core, b: in umbra in area with umbral dots, c: in umbra in area with large umbral dots / light bridge like structures, d: in umbra close to light bridge structure, e: in strong light bridge, f: in amorphous penumbral area, g: in penumbra.}
\label{fig:lamt}
\end{center}
\end{figure*}

The spectral evolution at different locations in the sunspot are shown in Figure~\ref{fig:lamt}. 
The spectral signature of dynamic fibrils is a diagonal line starting from maximum blueshift at the start of the event and with constant negative acceleration through 0~\kms\ to maximum redshift before disappearance. 
\citet{2008ApJ...673.1194L} 
showed similar spectral signatures for dynamic fibrils in Ca~II~8662 in their Figure~2. 
All $\lambda t$-diagrams show these signatures except panel \emph{a} which is from the dark umbral core dominated by umbral flashes. The regular appearance of the central reversals during the flashes is clearly visible.
Note that in the \Halpha\ line center images, the small dynamic fibrils seem to be present also in this region with \CaIIIR\ umbral flashes and absence of dynamic fibril spectral signatures.  
The amplitude of the dynamic fibril excursions is on the order $\pm10$~\kms. 
Note the small amplitudes in panel g, which is located in the penumbra in a region where the filaments are almost parallel to the disk center direction. 
This is where there is strongest signal in linear polarisation Stokes $Q$ and $U$ and the magnetic field is almost perpendicular to the line of sight. 
The plasma motion in the dynamic fibril is almost perpendicular to the line of sight in this region and therefore we observe only small amplitudes in the $\lambda t$-diagram. 

\begin{figure}[!t]
\begin{center}
\includegraphics[width=\columnwidth]{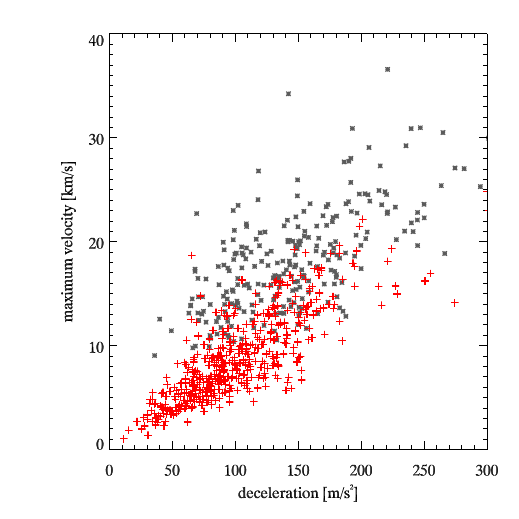}
\includegraphics[width=\columnwidth]{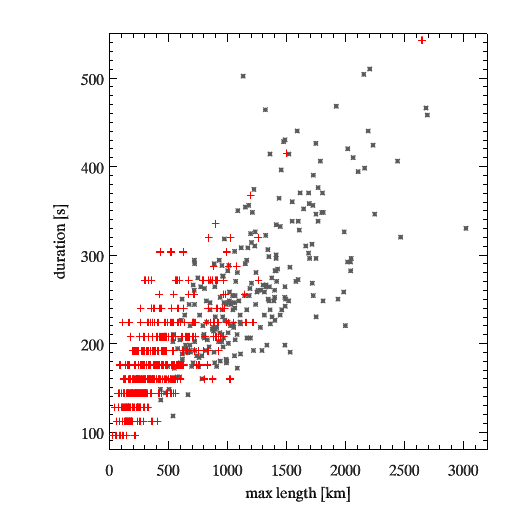}
\caption{Scatter plots of dynamic fibril properties. Red crosses are the measurements in the sunspot (460 individual dynamic fibrils), the smaller grey asterices are dynamic fibril properties from BDP07. 
}
\label{fig:scatt}
\end{center}
\end{figure}

A total of 460 
dynamic fibrils were measured by fitting parabolas to the path of their top in $xt$-diagrams.  
Values for maximum velocity, deceleration, duration and maximum length are presented in scatter plots in Figure~\ref{fig:scatt}. 
BDP07 
demonstrated that there is a clear correlation between maximum velocity and deceleration, and between duration and maximum length of dynamic fibrils in plage.
We show that dynamic fibrils in a sunspot show the same correlations, the data points of 
BDP07 
are added for comparison. 
Note that our data has significantly lower temporal resolution than the data of 
BDP07: 
almost 16 times lower. 
This explains the concentration of the duration measurements in our data in discrete bins that are integer multiples of the temporal cadence. 

\begin{figure*}[!t]
\begin{center}
\includegraphics[width=\textwidth]{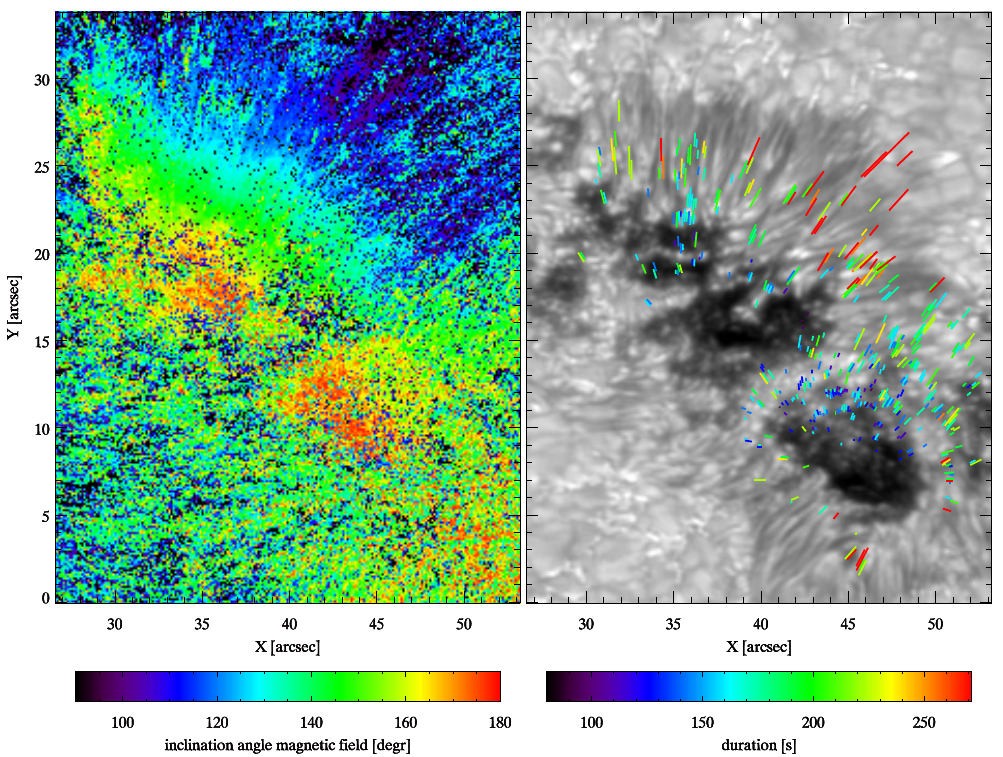}
\includegraphics[width=\columnwidth]{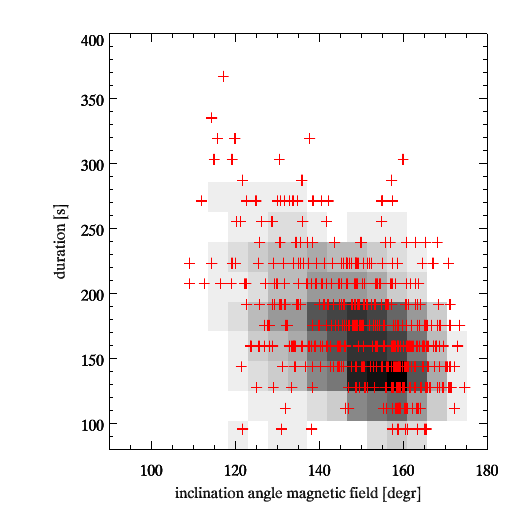}
\includegraphics[width=\columnwidth]{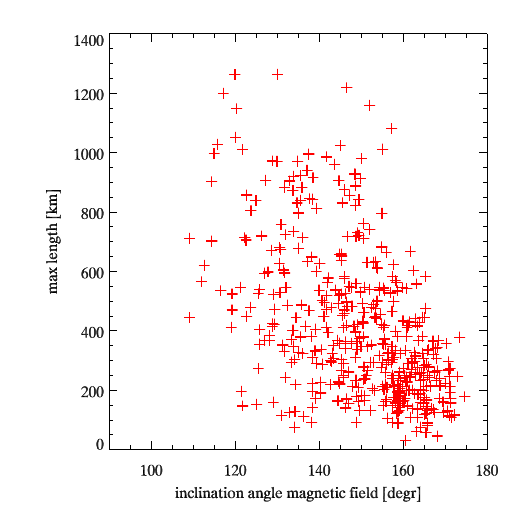}
\caption{Relation between magnetic field inclination and dynamic fibril duration and length. The top left panel shows a map of the magnetic field inclination from LTE inversion of the \CaIIIR\ Stokes data. 180\degr\ is vertical, negative polarity field, 90\degr\ is horizontal field. The top right panel shows a continuum \CaIIIR\ image with the path of the top of 460 dynamic fibrils drawn at their locations. The colour of the paths indicates the duration of the dynamic fibril. The bottom left panel shows a scatter plot of the duration of the dynamic fibrils against the inclination angle of the magnetic field at the lowest point of the path of the top of the dynamic fibril (red crosses). The inclination angle is the average value of a 3$\times$3 pixel area. In the background a smoothed density image of the same data points is shown with inverted colour table. The bottom right panel shows a scatter plot of the maximum length of the path of the top of the dynamic fibrils against the magnetic field inclination. }
\label{fig:binc}
\end{center}
\end{figure*}

The location of the 460 
analysed dynamic fibrils in the sunspot is illustrated by coloured lines in the right panel of Figure~\ref{fig:binc}.
The length of the lines indicates the maximum extension of the top of the dynamic fibrils, the colour is a measure of the duration. 
These properties can be compared to the inclination of the local magnetic field which can be inferred from the left panel. 
%
Despite the noise in the map, the general picture of the inclination in the lower chromosphere is that of almost vertical fields in the umbra and  progressively more horizontal fields in the penumbra at increasing distance from the umbra. 
%
This conforms to the paradigm of the general magnetic field topology in sunspots. 
The reliability of the magnetic field inclination measurements is obviously correlated with the signal in Stokes parameters $V$, and $Q$ and $U$. 
The strongest (relative) signals, particularly in linear polarisation, are found in the limb-side umbra and penumbra 
\citep[c.f. Figure 2 in][]{2013arXiv1304.0752D} 
and the dynamic fibrils are preferentially measured in these regions. 
From the comparison between the two panels in Figure~\ref{fig:binc}, a clear trend emerges for spatially longer, long-period dynamic fibrils to be located in regions with more horizontal magnetic fields. 

This correlation in duration versus magnetic field inclination angle and length versus inclination angle is clear from the scatter plots in the bottom panels of Figure~\ref{fig:binc}. 
The values for the field inclination angle are averages over 3$\times$3 pixels centred on the lowest point of the path of the top of the dynamic fibril. 
As mentioned above, the low temporal cadence of the data results in concentration of the duration values in discrete bins. 
This makes it difficult to differentiate between data points in high density areas in the left scatter plot. 
Therefore, a smoothed density map of the data points is shown in the background so that the visual trend in the data is better illustrated. 

\section{Discussion and conclusion}
\label{sec:discussion}

Our new
high-resolution SST observations reveal the ubiquitous presence of periodic jets of small extent in the chromosphere of sunspots. 
These jets follow parabolic trajectories in the space-time domain and appear to be the sunspot counterpart of the much longer dynamic fibrils found in plage regions,
of which the physics is well established
\citep[VH06, BDP07,][]{2007ApJ...666.1277H}. 
We find strong correlation between maximum velocity and deceleration, and between duration and maximum length for the tops of a sample of 460 
sunspot dynamic fibrils. 
These correlations agree well with corresponding measurements of dynamic fibrils in plage
\citep[VH06, BDP07,][]{2008ApJ...673.1194L, 
2008ApJ...673.1201L} 
and some mottles in quiet Sun \citep{2007ApJ...660L.169R}. 
In these studies, numerical simulations were used to identify these correlations as signatures that these periodic jets are driven by waves that propagate from the lower atmosphere into the chromosphere where they steepen into shocks. 
The sunspot dynamic fibrils have generally lower values for maximum length, duration, maximum velocity and deceleration than the dynamic fibrils in plage but given the convincing similarities we conclude that their driving mechanism and nature must be essentially the same. 
Therefore, this new solar feature represents a new class within this family of dynamic field-guided wave phenomena.

The more idealised (and therefore controlled) numerical simulations by 
\citet{2007ApJ...666.1277H} 
showed that there is a linear correlation between maximum velocity and deceleration for shock-wave driven chromospheric jets.
The slope of the correlation is basically determined by the period of the waves. 
Since the period of the wave train remains constant during the propagation and steepening into shock waves, there is a given time for which the amplitude can alter between extrema. 
So for fixed period waves of varying maximum velocity, the higher amplitude velocities will be subjected to larger deceleration. 
For varying periods, the longer period waves will experience lower deceleration.
These experiments suggest that the observed spread in the scatter plots of Figure~\ref{fig:scatt} can be partly attributed to variation in the excitation periods.
Spread in the scatter plot of duration versus maximum length is also introduced by projection effects in the measurement of the length.  
From the map of magnetic field inclination angle in Figure~\ref{fig:binc} it is clear that the viewing angles for dynamic fibrils in different areas in the sunspot varies significantly.
We do not attempt to correct for these projection effects but note that this effect introduces spread for the length measurements.
Obvious alternative major contributors to the spread are various errors in the measurements and observations. 

The simulations of 
\citet{2007ApJ...666.1277H} 
and later by 
\citet{2011ApJ...743..142H} 
elaborated on the idea that long period waves can propagate along inclined magnetic fields into higher and lower-density regions of the solar atmosphere 
(\citet{2004Natur.430..536D} 
and earlier investigated by 
\citet{1973SoPh...30...47M, 1977A&A....55..239B, 1990LNP...367..211S}). 
Inclined magnetic fields alter the acoustic cutoff period for waves that move along the magnetic field due to reduced effective gravity so that long-period waves become propagating into higher atmospheric regions where they otherwise would be evanescent in non-magnetic regions. 
From our observations, we can directly compare the inclination of the magnetic field \citep[converted to the local frame on the Sun as in ][]{2013scharmer} and the period of the sunspot dynamic fibrils (see Figure~\ref{fig:binc}). 
Indeed we find a clear trend of the longer duration dynamic fibrils being found in the regions with more inclined magnetic field.
This is direct observational support that this mechanism is at work for driving dynamic fibrils. 
VH06 and 
BDP07 
found indirect support from significant differences in dynamic fibril characteristics for two different regions where crude magnetic field extra-polations of low resolution MDI magnetic field maps indicated regional differences in the general magnetic field inclination. 
For our data, we can determine the local magnetic field vector from LTE inversions of the Stokes \CaIIIR\ data and 
measure the magnetic field inclination at approximately the same height as the dynamic fibrils. 
On much larger scales, the modification of the acoustic cutoff frequency by the change in the magnetic field inclination was demonstrated for a sunspot 
\citep{2006ApJ...647L..77M} 
and supergranular boundaries 
\citep{2006ApJ...648L.151J}. 

While we determine the magnetic field inclination at approximately the same height as the dynamic fibrils, it is the field inclination at lower heights that is important for the evanescence or propagation of waves to chromospheric heights.
If there is significant variation of the inclination with height, this would introduce spread in the scatter plots of Figure~\ref{fig:binc}  and reduce the correlation between the parameters. 
A possible way to study this effect would be to add photospheric spectropolarimetric diagnostics to the observing program but we remark that it would be extremely difficult to achieve sufficient resolution in the vertical domain (i.e., height in the sunspot atmosphere) in order to study this effect at high detail. 

Apart from the relation between magnetic field inclination and periods of the dynamic fibrils, the right panel of Figure~\ref{fig:binc} suggests a positive correlation between dynamic fibril length and duration.
A similar correlation was found for plage dynamic fibrils (BDP07) 
and in the MHD simulations of 
\citet{2009ApJ...701.1569M} 
and \citet{2011ApJ...743..142H}. 

The presence of upward propagating waves and shocks in sunspot atmospheres has been known from both observations
\citep[see e.g.,][]{1992sto..work..261L} 
and simulations
\citep[see e.g.,][]{2010ApJ...722..888B}. 
What is new from our high resolution SST observations is the high degree of fine structure in the form of small specs or fibrils that seems to be superposed on the large-scale oscillatory pattern
(i.e., the umbral flashes in \CaIIIR).
Possibly, there exist small-scale density fluctuations in the sunspot atmosphere resulting in local opacity differences that give rise to the apparent ejection of small-scale jets that are excited by larger-scale wave fronts. 
A high degree of fine-scale structuring in the sunspot photosphere has been known for a long time to exist in the form of for example umbral dots, light bridges and penumbral filaments 
\citep[for reviews see e.g.,][]{2003A&ARv..11..153S, 
2004ARA&A..42..517T}. 
The highest resolution sunspot observations suggest photospheric sub-structure down to the diffraction limit of present-day telescopes
\citep[see e.g., ][]{2002Natur.420..151S, 
2004SoPh..221...65L, 
2007Sci...318.1597I, 
thesis-vasco, 
2013A&A...557A...5H}. 
Recent simulation efforts explain the various fine-scale structures in sunspots as different manifestations of overturning convection in the strong magnetic field regime
\citep[see e.g., ][]{2006ApJ...641L..73S, 
2007ApJ...669.1390H, 
2011ApJ...729....5R}. 
It is likely that such fine-scale structuring in the photosphere extents to higher layers and also results in small-scale lateral variations in the vertical propagation of waves. 

The 2D simulations of 
\citet{2011ApJ...743..142H} 
showed that fine-scale structuring of chromospheric jets comes about naturally from slight differences in thermodynamic properties along neighboring field lines. 
Slight lateral differences lead to differences in shock speeds and thus the formation of a highly corrugated shock front surface.
The simulations show that initially smoothly and quasi-spherically expanding shock fronts nevertheless lead to skinny fingers of cold plasma protruding into the corona.
This seems to be a consequence of the fact that the path taken by the waves is not equally long for neighboring field lines and the speed of sound varies for each field line. 
Also, this process depends on the history of the plasma, i.e., the passage of previous waves (which lead to varying temperature structuring).
This gives an explanation why the fine structuring we see in the sunspot chromosphere in \Halpha\ and \CaIIIR\ is not a simple mapping of the photospheric structuring (in the form of umbral dots, light bridges and penumbral fine structure) but is also a result of factors that affect the wave propagation above the photosphere such as slight lateral thermodynamic differences and 
the local wave history.

It is plausible that the fine-scale structuring in Ca~II umbral flashes earlier reported from polarimetric inversions 
\citep{2000ApJ...544.1141S, 
2000Sci...288.1396S} 
and direct imaging
\citep{2009ApJ...696.1683S, 
2013A&A...557A...5H} 
is related to the small-scale sunspot dynamic fibrils we report here. 
%
It is however likely that they are different from the more transient micro-jets observed in Ca~II~H in the penumbra
\citep{2007Sci...318.1594K} 
and umbra
\citep{2013A&A...552L...1B}. 
These sunspot micro-jets appear to be a kind of jet that occurs much more infrequently, has shorter lifetime, is associated with brightening and appears to be only rising.
The penumbral micro-jets have been associated with magnetic reconnection processes
\citep[for example by][]{2008ApJ...686.1404R} 
and one might speculate that these could be the sunspot counterpart of the type ~II spicules at the limb.
Following that analogy, the sunspot dynamic fibrils we report here would be the sunspot counterparts of the wave-driven type~I spicules. 

Let us finish by adding some speculations on the general behavior of dynamic fibrils in different environments.
There are significant differences in maximum lengths if one compares dynamic fibrils in sunspots, plage, quiet Sun and coronal holes.
The longest dynamic fibrils are found in active region plage
(BDP07). 
In sunspots we observe a range of different lengths, with the shortest in the umbra, and progressively longer ones away from the umbral center and into the penumbra but all typically shorter than in plage. 
The quiet Sun equivalent of active region dynamic fibrils, or mottles, are also typically shorter than in plage 
\citep[cf.][]{2007ApJ...660L.169R}.  
At the limb, dynamic fibrils are observed as type I spicules and in the extensive study of 
\citet{2012ApJ...759...18P} 
only a very low number of parabolic spicules are found in quiet Sun and coronal holes, typically less than 3\% of the detected spicules, in some data sets even completely absent.
This apparent low abundance of type I spicules is probably caused by these structures being short and therefore hidden in the dense forrest of spicules at the limb.
In active regions on the other hand, the majority of detected spicules are of type I and reach maximum lengths that are similar to type II spicules in active regions, quiet Sun and coronal holes.
The variation in length for different regions is probably the result of a combination of factors: the input power or strength of the waves, the inclination angle of the magnetic field, and the spatial expansion of the magnetic field.
In quiet Sun and coronal holes there is relatively large wave power but significant field expansion from the photosphere to the chromosphere. 
This leads to the shock wave energy being spread out over a large area resulting in only short fibrils.
In sunspots, the wave power is much lower and there is relatively low field expansion. 
For the umbra, the combination of low wave power and almost vertical field leads to very short dynamic fibrils.
In addition, the magnetic field strength is strongest in the umbra leading to lower gas density and lower shock height.
Into the penumbra, the fields get more inclined leading to longer dynamic fibrils. 
There is probably relatively more convective input in the penumbra which leads to higher wave power and also contributes to longer dynamic fibrils. 
Finally, these factors conspire to the longest dynamic fibrils in plage: there is more convective energy input than in sunspots, inclined fields to enhance leakage of wave power into the chromosphere and not so much field expansion as in quiet Sun and coronal holes.

\acknowledgements
The authors thank Bart de Pontieu, Viggo Hansteen, Mats Carlsson and Rob Rutten for fruitful discussions.
The Swedish 1-m Solar Telescope was operated by the Institute for Solar Physics of the Royal Swedish Academy of Sciences in the Spanish Observatorio del Roque de los Muchachos of the Instituto de Astrof\'{\i}sica de Canarias.  
The authors gratefully acknowledge support from the
International Space Science Institute (ISSI).
Part of the computations were performed on resources
provided by the Swedish National Infrastructure for Computing (SNIC)
at Chalmers Centre for Computational Science and Engineering (C3SE) with project number SNIC002- 12-27.
This research was supported by the Research Council of Norway and by the European Research Council under the European Union's Seventh Framework Programme (FP7/2007-2013) / ERC Grant agreement nr. 291058.
This research has made use of NASA's Astrophysics Data System.


\begin{thebibliography}{55}
\expandafter\ifx\csname natexlab\endcsname\relax\def\natexlab#1{#1}\fi

\bibitem[{{Bard} \& {Carlsson}(2010)}]{2010ApJ...722..888B}
{Bard}, S. \& {Carlsson}, M. 2010, \apj, 722, 888

\bibitem[{{Bel} \& {Leroy}(1977)}]{1977A&A....55..239B}
{Bel}, N. \& {Leroy}, B. 1977, \aap, 55, 239

\bibitem[{{Bharti} {et~al.}(2013){Bharti}, {Hirzberger}, \&
  {Solanki}}]{2013A&A...552L...1B}
{Bharti}, L., {Hirzberger}, J., \& {Solanki}, S.~K. 2013, \aap, 552, L1

\bibitem[{{de la Cruz Rodr{\'{\i}}guez} {et~al.}(2013{\natexlab{a}}){de la Cruz
  Rodr{\'{\i}}guez}, {De Pontieu}, {Carlsson}, \& {Rouppe van der
  Voort}}]{2013ApJ...764L..11D}
{de la Cruz Rodr{\'{\i}}guez}, J., {De Pontieu}, B., {Carlsson}, M., \& {Rouppe
  van der Voort}, L.~H.~M. 2013{\natexlab{a}}, \apjl, 764, L11

\bibitem[{{de la Cruz Rodr{\'{\i}}guez} {et~al.}(2013{\natexlab{b}}){de la Cruz
  Rodr{\'{\i}}guez}, {Rouppe van der Voort}, {Socas-Navarro}, \& {van
  Noort}}]{2013arXiv1304.0752D}
{de la Cruz Rodr{\'{\i}}guez}, J., {Rouppe van der Voort}, L., {Socas-Navarro},
  H., \& {van Noort}, M. 2013{\natexlab{b}}, ArXiv e-prints 1304.0752

\bibitem[{{de la Cruz Rodr{\'{\i}}guez} {et~al.}(2012){de la Cruz
  Rodr{\'{\i}}guez}, {Socas-Navarro}, {Carlsson}, \&
  {Leenaarts}}]{2012delacruz}
{de la Cruz Rodr{\'{\i}}guez}, J., {Socas-Navarro}, H., {Carlsson}, M., \&
  {Leenaarts}, J. 2012, \aap, 543, A34

\bibitem[{{De Pontieu} {et~al.}(2004){De Pontieu}, {Erd{\'e}lyi}, \&
  {James}}]{2004Natur.430..536D}
{De Pontieu}, B., {Erd{\'e}lyi}, R., \& {James}, S.~P. 2004, \nat, 430, 536

\bibitem[{{De Pontieu} {et~al.}(2007{\natexlab{a}}){De Pontieu}, {Hansteen},
  {Rouppe van der Voort}, {van Noort}, \& {Carlsson}}]{2007ApJ...655..624D}
{De Pontieu}, B., {Hansteen}, V.~H., {Rouppe van der Voort}, L., {van Noort},
  M., \& {Carlsson}, M. 2007{\natexlab{a}}, \apj, 655, 624 (BDP07)

\bibitem[{{De Pontieu} {et~al.}(2007{\natexlab{b}}){De Pontieu}, {McIntosh},
  {Hansteen}, {Carlsson}, {Schrijver}, {Tarbell}, {Title}, {Shine}, {Suematsu},
  {Tsuneta}, {Katsukawa}, {Ichimoto}, {Shimizu}, \&
  {Nagata}}]{2007PASJ...59S.655D}
{De Pontieu}, B., {McIntosh}, S., {Hansteen}, V.~H., {et~al.}
  2007{\natexlab{b}}, \pasj, 59, 655

\bibitem[{{Hansteen} {et~al.}(2006){Hansteen}, {De Pontieu}, {Rouppe van der
  Voort}, {van Noort}, \& {Carlsson}}]{2006ApJ...647L..73H}
{Hansteen}, V.~H., {De Pontieu}, B., {Rouppe van der Voort}, L., {van Noort},
  M., \& {Carlsson}, M. 2006, \apjl, 647, L73 (VH06)

\bibitem[{{Heggland} {et~al.}(2007){Heggland}, {De Pontieu}, \&
  {Hansteen}}]{2007ApJ...666.1277H}
{Heggland}, L., {De Pontieu}, B., \& {Hansteen}, V.~H. 2007, \apj, 666, 1277

\bibitem[{{Heggland} {et~al.}(2011){Heggland}, {Hansteen}, {De Pontieu}, \&
  {Carlsson}}]{2011ApJ...743..142H}
{Heggland}, L., {Hansteen}, V.~H., {De Pontieu}, B., \& {Carlsson}, M. 2011,
  \apj, 743, 142

\bibitem[{{Heinemann} {et~al.}(2007){Heinemann}, {Nordlund}, {Scharmer}, \&
  {Spruit}}]{2007ApJ...669.1390H}
{Heinemann}, T., {Nordlund}, {\AA}., {Scharmer}, G.~B., \& {Spruit}, H.~C.
  2007, \apj, 669, 1390

\bibitem[{{Henriques}(2012)}]{2012A&A...548A.114H}
{Henriques}, V.~M.~J. 2012, \aap, 548, A114

\bibitem[{Henriques(2013)}]{thesis-vasco}
Henriques, V. M.~J. 2013, PhD thesis, Stockholm University,
  \url{http://urn.kb.se/resolve?urn=urn:nbn:se:su:diva-86798}

\bibitem[{{Henriques} \& {Kiselman}(2013)}]{2013A&A...557A...5H}
{Henriques}, V.~M.~J. \& {Kiselman}, D. 2013, \aap, 557, A5

\bibitem[{{Ichimoto} {et~al.}(2007){Ichimoto}, {Suematsu}, {Tsuneta},
  {Katsukawa}, {Shimizu}, {Shine}, {Tarbell}, {Title}, {Lites}, {Kubo}, \&
  {Nagata}}]{2007Sci...318.1597I}
{Ichimoto}, K., {Suematsu}, Y., {Tsuneta}, S., {et~al.} 2007, Science, 318,
  1597

\bibitem[{{Jefferies} {et~al.}(2006){Jefferies}, {McIntosh}, {Armstrong},
  {Bogdan}, {Cacciani}, \& {Fleck}}]{2006ApJ...648L.151J}
{Jefferies}, S.~M., {McIntosh}, S.~W., {Armstrong}, J.~D., {et~al.} 2006,
  \apjl, 648, L151

\bibitem[{{Katsukawa} {et~al.}(2007){Katsukawa}, {Berger}, {Ichimoto}, {Lites},
  {Nagata}, {Shimizu}, {Shine}, {Suematsu}, {Tarbell}, {Title}, \&
  {Tsuneta}}]{2007Sci...318.1594K}
{Katsukawa}, Y., {Berger}, T.~E., {Ichimoto}, K., {et~al.} 2007, Science, 318,
  1594

\bibitem[{{Langangen} {et~al.}(2008{\natexlab{a}}){Langangen}, {Carlsson},
  {Rouppe van der Voort}, {Hansteen}, \& {De Pontieu}}]{2008ApJ...673.1194L}
{Langangen}, {\O}., {Carlsson}, M., {Rouppe van der Voort}, L., {Hansteen}, V.,
  \& {De Pontieu}, B. 2008{\natexlab{a}}, \apj, 673, 1194

\bibitem[{{Langangen} {et~al.}(2008{\natexlab{b}}){Langangen}, {Rouppe van der
  Voort}, \& {Lin}}]{2008ApJ...673.1201L}
{Langangen}, {\O}., {Rouppe van der Voort}, L., \& {Lin}, Y.
  2008{\natexlab{b}}, \apj, 673, 1201

\bibitem[{{Lites}(1992)}]{1992sto..work..261L}
{Lites}, B.~W. 1992, in NATO ASIC Proc. 375: Sunspots. Theory and Observations,
  ed. J.~H. {Thomas} \& N.~O. {Weiss}, 261--302

\bibitem[{{Lites} {et~al.}(2004){Lites}, {Scharmer}, {Berger}, \&
  {Title}}]{2004SoPh..221...65L}
{Lites}, B.~W., {Scharmer}, G.~B., {Berger}, T.~E., \& {Title}, A.~M. 2004,
  \solphys, 221, 65

\bibitem[{{Mart{\'{\i}}nez-Sykora} {et~al.}(2009){Mart{\'{\i}}nez-Sykora},
  {Hansteen}, {De Pontieu}, \& {Carlsson}}]{2009ApJ...701.1569M}
{Mart{\'{\i}}nez-Sykora}, J., {Hansteen}, V., {De Pontieu}, B., \& {Carlsson},
  M. 2009, \apj, 701, 1569

\bibitem[{{McIntosh} \& {Jefferies}(2006)}]{2006ApJ...647L..77M}
{McIntosh}, S.~W. \& {Jefferies}, S.~M. 2006, \apjl, 647, L77

\bibitem[{{Michalitsanos}(1973)}]{1973SoPh...30...47M}
{Michalitsanos}, A.~G. 1973, \solphys, 30, 47

\bibitem[{{Nagashima} {et~al.}(2007){Nagashima}, {Sekii}, {Kosovichev},
  {Shibahashi}, {Tsuneta}, {Ichimoto}, {Katsukawa}, {Lites}, {Nagata},
  {Shimizu}, {Shine}, {Suematsu}, {Tarbell}, \& {Title}}]{2007PASJ...59S.631N}
{Nagashima}, K., {Sekii}, T., {Kosovichev}, A.~G., {et~al.} 2007, \pasj, 59,
  631

\bibitem[{{Pereira} {et~al.}(2012){Pereira}, {De Pontieu}, \&
  {Carlsson}}]{2012ApJ...759...18P}
{Pereira}, T.~M.~D., {De Pontieu}, B., \& {Carlsson}, M. 2012, \apj, 759, 18

\bibitem[{{Rempel}(2011)}]{2011ApJ...729....5R}
{Rempel}, M. 2011, \apj, 729, 5

\bibitem[{{Rouppe van der Voort} {et~al.}(2007){Rouppe van der Voort}, {De
  Pontieu}, {Hansteen}, {Carlsson}, \& {van Noort}}]{2007ApJ...660L.169R}
{Rouppe van der Voort}, L.~H.~M., {De Pontieu}, B., {Hansteen}, V.~H.,
  {Carlsson}, M., \& {van Noort}, M. 2007, \apjl, 660, L169

\bibitem[{{Rouppe van der Voort} {et~al.}(2003){Rouppe van der Voort},
  {Rutten}, {S{\"u}tterlin}, {Sloover}, \& {Krijger}}]{2003A&A...403..277R}
{Rouppe van der Voort}, L.~H.~M., {Rutten}, R.~J., {S{\"u}tterlin}, P.,
  {Sloover}, P.~J., \& {Krijger}, J.~M. 2003, \aap, 403, 277

\bibitem[{{Ryutova} {et~al.}(2008){Ryutova}, {Berger}, {Frank}, \&
  {Title}}]{2008ApJ...686.1404R}
{Ryutova}, M., {Berger}, T., {Frank}, Z., \& {Title}, A. 2008, \apj, 686, 1404

\bibitem[{{Scharmer} {et~al.}(2003{\natexlab{a}}){Scharmer}, {Bjelksj{\"o}},
  {Korhonen}, {Lindberg}, \& {Petterson}}]{2003SPIE.4853..341S}
{Scharmer}, G.~B., {Bjelksj{\"o}}, K., {Korhonen}, T.~K., {Lindberg}, B., \&
  {Petterson}, B. 2003{\natexlab{a}}, in Society of Photo-Optical
  Instrumentation Engineers (SPIE) Conference Series, Vol. 4853, Society of
  Photo-Optical Instrumentation Engineers (SPIE) Conference Series, ed. S.~L.
  {Keil} \& S.~V. {Avakyan}, 341--350

\bibitem[{{Scharmer} {et~al.}(2013){Scharmer}, {de la Cruz Rodriguez},
  {S{\"u}tterlin}, \& {Henriques}}]{2013scharmer}
{Scharmer}, G.~B., {de la Cruz Rodriguez}, J., {S{\"u}tterlin}, P., \&
  {Henriques}, V.~M.~J. 2013, \aap, 553, A63

\bibitem[{{Scharmer} {et~al.}(2003{\natexlab{b}}){Scharmer}, {Dettori},
  {Lofdahl}, \& {Shand}}]{2003SPIE.4853..370S}
{Scharmer}, G.~B., {Dettori}, P.~M., {Lofdahl}, M.~G., \& {Shand}, M.
  2003{\natexlab{b}}, in Society of Photo-Optical Instrumentation Engineers
  (SPIE) Conference Series, Vol. 4853, Society of Photo-Optical Instrumentation
  Engineers (SPIE) Conference Series, ed. {S.~L.~Keil \& S.~V.~Avakyan},
  370--380

\bibitem[{{Scharmer} {et~al.}(2002){Scharmer}, {Gudiksen}, {Kiselman},
  {L{\"o}fdahl}, \& {Rouppe van der Voort}}]{2002Natur.420..151S}
{Scharmer}, G.~B., {Gudiksen}, B.~V., {Kiselman}, D., {L{\"o}fdahl}, M.~G., \&
  {Rouppe van der Voort}, L.~H.~M. 2002, \nat, 420, 151

\bibitem[{{Scharmer} {et~al.}(2008){Scharmer}, {Narayan}, {Hillberg}, {de la
  Cruz Rodr{\'{\i}}guez}, {L{\"o}fdahl}, {Kiselman}, {S{\"u}tterlin}, {van
  Noort}, \& {Lagg}}]{2008ApJ...689L..69S}
{Scharmer}, G.~B., {Narayan}, G., {Hillberg}, T., {et~al.} 2008, \apjl, 689,
  L69

\bibitem[{{Schnerr} {et~al.}(2011){Schnerr}, {de La Cruz Rodr{\'{\i}}guez}, \&
  {van Noort}}]{2011A&A...534A..45S}
{Schnerr}, R.~S., {de La Cruz Rodr{\'{\i}}guez}, J., \& {van Noort}, M. 2011,
  \aap, 534, A45

\bibitem[{{Sch{\"u}ssler} \& {V{\"o}gler}(2006)}]{2006ApJ...641L..73S}
{Sch{\"u}ssler}, M. \& {V{\"o}gler}, A. 2006, \apjl, 641, L73

\bibitem[{{Selbing}(2010)}]{2010selbing}
{Selbing}, J. 2010, ArXiv e-prints: 1010.4142

\bibitem[{{Shibata} \& {Suematsu}(1982)}]{1982SoPh...78..333S}
{Shibata}, K. \& {Suematsu}, Y. 1982, \solphys, 78, 333

\bibitem[{{Shine} {et~al.}(1994){Shine}, {Title}, {Tarbell}, {Smith}, {Frank},
  \& {Scharmer}}]{1994shine}
{Shine}, R.~A., {Title}, A.~M., {Tarbell}, T.~D., {et~al.} 1994, \apj, 430, 413

\bibitem[{{Socas-Navarro} {et~al.}(2009){Socas-Navarro}, {McIntosh}, {Centeno},
  {de Wijn}, \& {Lites}}]{2009ApJ...696.1683S}
{Socas-Navarro}, H., {McIntosh}, S.~W., {Centeno}, R., {de Wijn}, A.~G., \&
  {Lites}, B.~W. 2009, \apj, 696, 1683

\bibitem[{{Socas-Navarro} {et~al.}(2000{\natexlab{a}}){Socas-Navarro},
  {Trujillo Bueno}, \& {Ruiz Cobo}}]{2000ApJ...544.1141S}
{Socas-Navarro}, H., {Trujillo Bueno}, J., \& {Ruiz Cobo}, B.
  2000{\natexlab{a}}, \apj, 544, 1141

\bibitem[{{Socas-Navarro} {et~al.}(2000{\natexlab{b}}){Socas-Navarro},
  {Trujillo Bueno}, \& {Ruiz Cobo}}]{2000Sci...288.1396S}
{Socas-Navarro}, H., {Trujillo Bueno}, J., \& {Ruiz Cobo}, B.
  2000{\natexlab{b}}, Science, 288, 1396

\bibitem[{{Socas-Navarro} {et~al.}(2000{\natexlab{c}}){Socas-Navarro},
  {Trujillo Bueno}, \& {Ruiz Cobo}}]{2000socas-navarro}
{Socas-Navarro}, H., {Trujillo Bueno}, J., \& {Ruiz Cobo}, B.
  2000{\natexlab{c}}, \apj, 530, 977

\bibitem[{{Solanki}(2003)}]{2003A&ARv..11..153S}
{Solanki}, S.~K. 2003, \aapr, 11, 153

\bibitem[{{Suematsu}(1990)}]{1990LNP...367..211S}
{Suematsu}, Y. 1990, in Lecture Notes in Physics, Berlin Springer Verlag, Vol.
  367, Progress of Seismology of the Sun and Stars, ed. Y.~{Osaki} \&
  H.~{Shibahashi}, 211

\bibitem[{{Suematsu} {et~al.}(1982){Suematsu}, {Shibata}, {Neshikawa}, \&
  {Kitai}}]{1982SoPh...75...99S}
{Suematsu}, Y., {Shibata}, K., {Neshikawa}, T., \& {Kitai}, R. 1982, \solphys,
  75, 99

\bibitem[{{Thomas} \& {Weiss}(2004)}]{2004ARA&A..42..517T}
{Thomas}, J.~H. \& {Weiss}, N.~O. 2004, \araa, 42, 517

\bibitem[{{Title} \& {Rosenberg}(1981)}]{title81SOUP}
{Title}, A. \& {Rosenberg}, W. 1981, Optical Engineering, 20, 815

\bibitem[{{Tsiropoula} {et~al.}(2012){Tsiropoula}, {Tziotziou}, {Kontogiannis},
  {Madjarska}, {Doyle}, \& {Suematsu}}]{2012SSRv..169..181T}
{Tsiropoula}, G., {Tziotziou}, K., {Kontogiannis}, I., {et~al.} 2012, \ssr,
  169, 181

\bibitem[{{van Noort} {et~al.}(2005){van Noort}, {Rouppe van der Voort}, \&
  {L{\"o}fdahl}}]{2005SoPh..228..191V}
{van Noort}, M., {Rouppe van der Voort}, L., \& {L{\"o}fdahl}, M.~G. 2005,
  \solphys, 228, 191

\bibitem[{{van Noort} \& {Rouppe van der Voort}(2008)}]{2008A&A...489..429V}
{van Noort}, M.~J. \& {Rouppe van der Voort}, L.~H.~M. 2008, \aap, 489, 429

\bibitem[{{Vissers} \& {Rouppe van der Voort}(2012)}]{2012ApJ...750...22V}
{Vissers}, G. \& {Rouppe van der Voort}, L. 2012, \apj, 750, 22

\end{thebibliography}

\end{document}